\documentclass[twocolumn,twocolappendix,
numberedappendix,apj]{openjournal}

\usepackage{newtxtext,newtxmath}
\usepackage[T1]{fontenc}
\usepackage{natbib}
\usepackage{graphicx}
\usepackage{amsmath,amssymb,amstext}
\usepackage{physics}
\usepackage{url}
\usepackage[colorlinks=true,breaklinks,
citecolor=blue,urlcolor=blue]{hyperref}
\usepackage{xspace}

\usepackage{array}
\newcolumntype{P}[1]{>{\centering\arraybackslash}p{#1}}
\newcolumntype{M}[1]{>{\centering\arraybackslash}m{#1}}

\newcommand{\hMpc}{\,h^{-1}\;{\rm Mpc}}
\newcommand{\hMpcdens}{\,h^3\;{\rm Mpc}^{-3}}
\newcommand{\hGpc}{\,h^{-1}\;{\rm Gpc}}
\newcommand{\hMsun}{\,h^{-1}\, M_\sun}
\newcommand{\Quijote}{\textsc{Quijote}\xspace}
\newcommand{\Gudhi}{\textsc{Gudhi}\xspace}

\begin{document}
\journalinfo{The Open Journal of Astrophysics}

\title{Cosmological information content of Betti curves and $k$-nearest neighbor distributions\vspace{-1.25cm}}
\shorttitle{Cosmology with Betti curves and $k$NNs}

\author{Aaron Ouellette$^{1,2,3\ast}$ and Gilbert Holder$^{1,2}$}
\shortauthors{Ouellette \& Holder}

\affiliation{
    $^1$ Department of Physics, University of Illinois Urbana-Champaign, Urbana, IL, 61801, USA
}
\affiliation{
    $^2$ Illinois Center for Advanced Studies of the Universe, University of Illinois Urbana-Champaign, Urbana, IL, 61801, USA
}
\affiliation{
    $^3$ Center for AstroPhysical Surveys, National Center for Supercomputing Applications, Urbana, IL, 61801, USA
}
\thanks{$^\ast$\href{mailto:aaronjo2@illinois.edu}{aaronjo2@illinois.edu}}

\begin{abstract}
    We compare the cosmological constraints that can be obtained from halo clustering on non-linear scales ($2 \hMpc < r < 50 \hMpc$) using Betti curves, a topological summary statistic, and $k$-th nearest neighbor ($k$NN) distributions. We quantify the information content of each summary statistic through Fisher matrices computed from the \Quijote simulations. Due to the use of simulation-based Fisher forecasts, we pay careful attention to the convergence of the Fisher matrices by looking at their eigendecompositions. We find that, in general, only two directions in the parameter space have constraints that are well converged given the number of \Quijote simulations available. We then compare the information content of each summary statistic in the reduced parameter space $\{\Omega_m, \sigma_8\}$. We find that almost all of the information present in the Betti curves comes from the first two, $\beta_0$ and $\beta_1$, which track the number of connected components and one-dimensional loops respectively, and almost no constraining power comes from $\beta_2$ which tracks the number of topological voids. In comparison, we find that the $k$NNs provide very competitive constraints along with several potential advantages in regards to real data. Finally, we find that while the $k$NNs and Betti curves provide some complementary constraints, they are not fully independent, potentially indicating a connection between the two statistics.
\end{abstract}

%\keywords{cosmological parameters -- large-scale structure of Universe}

\section{Introduction}

In recent years significant effort has been spent exploring new summary statistics to characterize the large-scale structure of matter in the Universe. While the two-point correlation function, whether in real or Fourier space (as the power spectrum), remains the standard choice for state-of-the-art cosmological analyses \citep[see, for example,][]{Abbott2022,DESI_Collaboration2024} due to well-developed modeling and measurement pipelines, it is well known that the density field becomes highly non-Gaussian through gravitational evolution causing the two-point function to no longer be effective at extracting information from small scales. This motivates the search for new summary statistics that can effectively extract information from the non-linear small scales. A few examples of these beyond-two-point summary statistics include: the bispectrum and other higher $n$-point functions \citep{Sefusatti2006,Hahn2020}, one-point PDFs \citep{Uhlemann2020}, $k$-nearest neighbor ($k$NN) distance distributions \citep{Banerjee2021,Yuan2023}, various topological summary statistics \citep{Weygaert2011,Biagetti2021,CisewskiKehe2022,Ouellette2023,Yip2024}, wavelet scattering transforms \citep{Valogiannis2022}, various extrema statistics \citep{Wang2025}, as well as many others. 

With the large number of existing and seemingly disparate summary statistics, it is potentially useful to study the possible connections between them. Here, we focus on Betti curves, a topological summary statistic based on persistent homology, and $k$NN distributions which encode information from all levels of the $n$-point hierarchy. Betti curves have previously been shown to encode non-Gaussian information present in cosmological fields \citep{Ouellette2023,Bermejo2024} and provide a way of rigorously defining intuitive concepts such as loops and voids. $k$NNs, in particular, seem to provide a promising way to establish connections between different probes. They provide efficient summaries of non-Gaussian information, while also being directly connected to the 2-point function \citep{Banerjee2021,Yuan2023}. They can be used to measure the clustering of discrete datasets \citep{Wang2022} as well as continuous ones \citep{Banerjee2023} and can be used to measure cross-correlations \citep{Banerjee2021b,Zhou2024,Gupta2024}. $k$NNs have also been shown to be related to the void probability function \citep{Banerjee2021}, the counts-in-cells summary statistic \citep{Coulton2024}, and Minkowski functionals \citep{Gangopadhyay2025}. 

In this paper, we aim to do a consistent comparison between the amount of cosmological information that can be extracted from halo catalogs using Betti curves and $k$NN distributions. We do this using Fisher matrices computed from cosmological simulations, but we pay extra attention to the issue of convergence and propose a new simple method to robustly measure the convergence of Fisher matrices.

Our paper is structured as follows: in Sections~\ref{sec:bc} and \ref{sec:knn} we provide brief overviews of Betti curves and $k$NN distributions. We provide an overview of the simulated catalogs and the methods used to compute the summary statistics in Sections~\ref{sec:sim} and \ref{sec:stat}. In Section~\ref{sec:fisher} we outline the Fisher matrix formalism and our analysis methods. In Section~\ref{sec:res} we explore the resulting cosmological constraints obtained from the considered summary statistics. Finally, in Section~\ref{sec:concl}, we discuss our findings and conclude.

\section{Betti curves and persistent homology}
\label{sec:bc}

Computing topological summaries for a point cloud dataset (potentially representing the positions of halos or galaxies) relies on persistent homology \citep{Edelsbrunner2002,Carlsson2005,Carlsson2009}, a formalism that allows one to characterize the topological properties of a dataset as a function of scale. Here we provide a brief overview of the necessary concepts and terminology. See \cite{Otter2017} for a more detailed overview.

For a point cloud dataset, a topological space can be defined by constructing a \textit{simplicial complex} which is simply an assembly of $k$-simplices that defines how points in the dataset are connected. In three dimensions, the relevant simplices are vertices ($k=0$), edges ($k=1$), triangles ($k=2$), and tetrahedra ($k=3$). Any simplicial complex must also be closed under the intersection of simplices and under taking faces, i.e., for any two simplices in the complex, their intersection must also be in the complex and for any simplex in the complex, its face must also be in the complex. The topological properties of a simplicial complex are given by its \textit{homology groups} $H_d$ for $d \in \{0, 1, 2, ...\}$. The elements of $H_d$ are equivalence classes of $d$-cycles, or more simply, independent ``holes" of dimension $d$. The relevant cycles in three dimensions are connected components ($d=0$), loops ($d=1$), and cavities ($d=2$). Finally, in order to associate topological spaces to different scales in the dataset, we must define a \textit{filtration} which is simply a nested family of simplicial complexes where each subcomplex (a subset of a simplicial complex that itself is a simplicial complex) corresponds to some value of a filtration parameter. Persistent homology allows one to compute the full evolution of the homology groups as a function of the filtration parameter for arbitrary datasets.

\begin{figure}
    \centering
    \includegraphics[width=\linewidth]{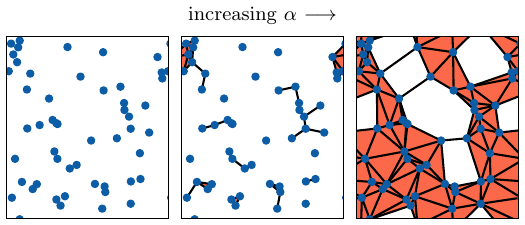}
    \caption{An example of an $\alpha$-complex filtration using random points in a 2D periodic box. As the filtration parameter $\alpha$ increases, larger simplices are added to the complex and more neighboring points get connected. Topological features such as loops appear in the complex at some value of $\alpha$ and then disappear at a larger value as more simplices are added. Betti curves ($\beta_0$ and $\beta_1$ in 2D) track the numbers of these features as a function of $\alpha$. In the first panel ($\alpha = 0$), the complex consists only of disconnected points, and $\beta_0$ is at its maximum value and $\beta_1 = 0$. In the middle panel, many connections have formed, causing $\beta_0$ to decrease, but no loops exist yet. In the last panel, everything is connected ($\beta_0 = 1$) and several loops have formed ($\beta_1 = 6$). If $\alpha$ is increased further, these loops will get filled in, resulting in an object with a trivial topology ($\beta_0 = 1$, $\beta_1 = 0$).}
    \label{fig:ex}
\end{figure}

The above description is very general and abstract, because the specific definition of how a simplicial complex is constructed and filtered is arbitrary and many different variants exist in the literature. In this work, as in \cite{Ouellette2023}, we focus on the $\alpha$-complex filtration \citep{Edelsbrunner1983,Edelsbrunner1994}, which is defined in terms of the Delaunay triangulation of the points in the dataset. For all values of the filtration parameter $\alpha \ge 0$, the $\alpha$-complex is a subcomplex of the Delaunay triangulation. At $\alpha = 0$, the complex consists only of vertices given by the points in the original dataset. For $\alpha > 0$, simplices from the Delaunay triangulation are added to the complex if their circumsphere has a radius $\le \alpha$. To make these ideas more concrete, we provide a visualization of a 2D $\alpha$-complex filtration in Figure~\ref{fig:ex}.

There are many different ways to construct summary statistics based on the topological features identified through persistent homology \cite[see for example,][]{Berry2020}. Here, we focus on Betti curves -- a functional summary that simply counts the number of homology classes of a given dimension as a function of scale. The Betti curve $\beta_d(r)$ represents the number of homology classes of dimension $d$ present at the filtration scale $\alpha = r$. Betti curves only contain a subset of the information obtained through persistent homology, but are interpretable and can easily be used for statistical inference. Betti curves have recently been used for a variety of applications in cosmology \citep{Pranav2017,Feldbrugge2019,Wilding2021,Elbers2019,Elbers2023,Giri2021,Tymchyshyn2023,Jalali2024}.

\section{Nearest neighbor distribution functions}
\label{sec:knn}

$k$NN statistics summarize the distributions of distances from volume-filling points to the $k$-th nearest data point. The $k=1$ case was first used by \cite{Ryden1984} to characterize voids in galaxy catalogs, while the more general statistic for arbitrary $k \ge 1$ was introduced as a cosmological probe by \cite{Banerjee2021}. These distributions are usually expressed in terms of their cumulative distribution functions (CDFs): CDF$_{k\text{NN}}(r)$ is the fraction of volume-filling points with distance to the $k$-th nearest data point $< r$. \cite{Banerjee2021} showed that the full set of $k$NN CDFs for all values of $k$ is sensitive to the full hierarchy of $n$-point functions. The significant advantage of $k$NNs is that their computational cost is constant for all values of $k$, in contrast to the exponential scaling of the $n$-point functions.

The $k$NNs can also be defined by calculating distances between pairs of data points instead of data-random pairs \citep{Yuan2023}. Throughout this paper, we will label the $k$NNs utilizing volume filling points as DR-$k$NNs and the $k$NNs utilizing pairs of data points as DD-$k$NNs. While DR- and DD-$k$NNs are defined in very similar ways, they are sensitive to the $n$-point correlation functions in different ways, allowing them to extract complementary information. Specifically, the DR-$k$NNs tend to probe low-density regions, while the DD-$k$NNs tend to probe high-density regions \citep{Yuan2023}.

\section{Simulations and halo catalogs}
\label{sec:sim}
We use the \Quijote suite of cosmological simulations \citep{Villaescusa-Navarro2020} to compare the sensitivity of Betti curves and $k$NNs to variations in cosmological parameters. \Quijote is a large set of N-body simulations covering a wide cosmological parameter space in the $\nu w$CDM model, each of which follows the evolution of $512^3$ dark matter particles in a $1 \hGpc$ box. \Quijote contains a large number of simulations at a fiducial cosmology, given by $\Omega_m = 0.3175$, $\Omega_b = 0.049$, $\sigma_8 = 0.834$, $n_s = 0.9624$, $h = 0.6711$, $M_\nu = 0$ eV, and $w = -1$, to enable the calculation of covariance matrices for arbitrary summary statistics. \Quijote also contains a large set of simulations that perturb each of these parameters one at a time in order to enable to calculation of numerical derivatives with respect to the cosmological parameters. Each simulation has a publicly available halo catalog which was generated using the Friends-of-Friends algorithm with a linking parameter of $b=0.2$.

In this analysis we focus only on the redshift $z=0$ halo catalogs and consider the reduced parameter space $\{\Omega_m, \sigma_8, h, n_s, M_\nu\}$. Since both the Betti curves and $k$NNs are very sensitive to the number density of the data points \citep{Ouellette2023,Banerjee2021}, we keep the number density of halos ($\bar{n}$) constant by selecting the top $150,000$ most massive halos from each simulation ($\bar{n} = 1.5 \times 10^{-4} \hMpcdens$). We use this number density cut instead of a halo mass cut to ensure that the Betti curves and $k$NNs do not mainly respond to changes in the number of halos but to changes in the halo clustering. The chosen number density cut corresponds roughly to a mass cut of $3.3 \times 10^{13} \hMsun$ in the fiducial cosmology.

\section{Summary statistics}
\label{sec:stat}
\subsection{2-point correlation function}
As a baseline comparison, we compute the 2-point correlation functions $\xi(r)$ for each halo sample using the \textsc{Corrfunc}\footnote{\url{https://corrfunc.readthedocs.io/en/master/}} library \citep{Sinha2019,Sinha2020}. We compute $\xi(r)$ using 32 logarithmic bins between $2 \hMpc < r < 50 \hMpc$. 

\subsection{Betti curves}
We calculate the Betti curves for each halo sample using the same steps outlined in \cite{Ouellette2023}. Briefly, we use the \Gudhi \footnote{\url{https://gudhi.inria.fr/}} library to construct $\alpha$-complex triangulations on top of each halo sample in a 3D periodic box and then calculate the Betti curves as a function of filtration scale $\alpha$. We normalize the Betti curves by the total number of halos in the sample resulting in a measure of the number of homology classes per halo. 

\begin{figure}
    \centering
    \includegraphics[width=0.9\linewidth]{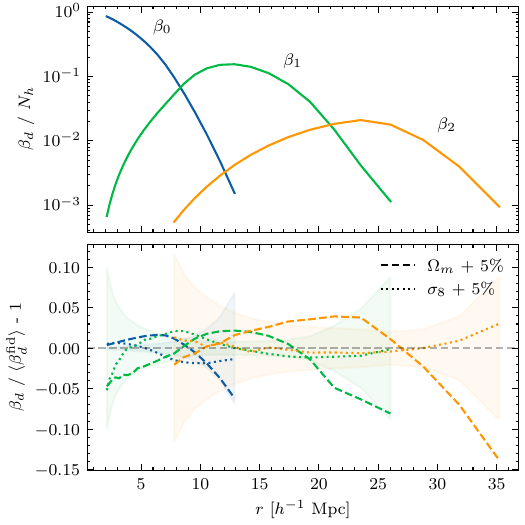}
    \caption{Top panel: Average halo Betti curves computed from 5,000 simulations at the fiducial cosmology. The Betti curves are normalized by the number of halos in the catalog ($N_h = 150,000$). Bottom panel: Fractional variations in the Betti curves when changing $\Omega_m$ / $\sigma_8$ by 5\%. The faint shaded regions indicate the range of $1\sigma$ variations at the fiducial cosmology.}
    \label{fig:betti}
\end{figure}

\begin{figure}
    \centering
    \includegraphics[width=\linewidth]{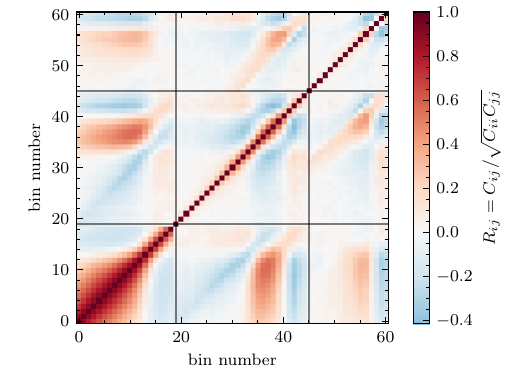}
    \caption{Betti curve correlation matrix computed from 5,000 simulations at the fiducial cosmology. The three sub-matrices along the diagonal represent the correlation matrices for $\beta_0$, $\beta_1$, and $\beta_2$, in order of increasing bin number.}
    \label{fig:betti_cov}
\end{figure}

We apply the same scale cuts as above, $2 \hMpc < r < 50 \hMpc$, to the Betti curves, but we additionally require that $\beta_d > 5\times10^{-4}$ for each of the Betti curves $\beta_d$. This ensures that we do not go too far into the tails of the Betti curves and that the summary statistics remain well-described by a multivariate Gaussian distribution (see Appendix~\ref{app:gauss}). Finally, to keep the analysis roughly consistent across summary statistics, we evaluate each Betti curve at the same number of points per logarithmic interval in $r$ as for the 2-point correlation function. We show the resulting halo Betti curves evaluated at the fiducial cosmology in Figure~\ref{fig:betti} and the full correlation matrix for the Betti curves in Figure~\ref{fig:betti_cov}.

\subsection{$k$NN CDFs}
To calculate the $k$NN CDFs for each halo sample, we use the $k$-d tree as implemented by \textsc{scipy} that allows efficient querying of the distances from a set of points to their $k$ nearest neighbors. We compute both the DR-$k$NN CDFs and the DD-$k$NN CDFs. For the DR-$k$NNs, $k$-th nearest neighbor distances are calculated between a set of $10^6$ uniform random points that fill the simulation volume and the halo sample, while for the DD-$k$NNs, the distances are calculated between points in the halo sample. We then construct the empirical CDF for either the DR-$k$NN or the DD-$k$NN once all of the nearest neighbor distances have been calculated for a given value of $k$. In this analysis we use $k = \{1, 2, 4, 8, 16\}$. 

For better visualization of the distribution tails, it is common to plot the peaked $k$NN CDFs, which is defined as
\begin{equation}
    \text{pCDF}_{k\text{NN}}(r) = 
    \begin{cases}
        \text{CDF}_{k\text{NN}}(r) & \text{CDF}_{k\text{NN}}(r) < 0.5 \\
        1 - \text{CDF}_{k\text{NN}}(r) & \text{CDF}_{k\text{NN}}(r) \ge 0.5.
    \end{cases}
\end{equation}

\begin{figure}
    \centering
    \includegraphics[width=0.9\linewidth]{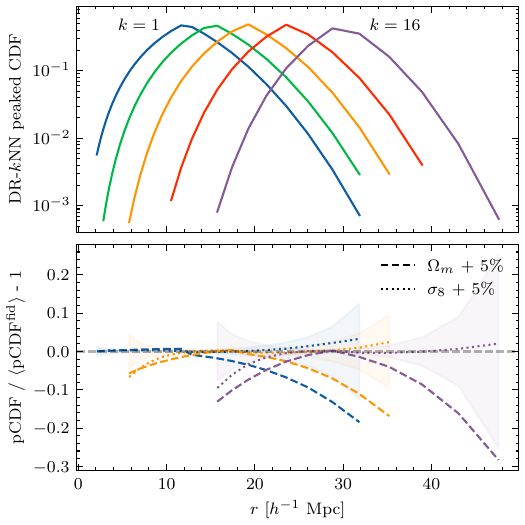}
    \caption{Top panel: Average DR-$k$NNs for the fiducial cosmology. For visualization purposes, we plot the peaked CDF. From left to right, the curves represent the $k$NN CDFs for $k = 1, 2, 4, 8, 16$. Bottom panel: Fractional variations in the $k=1$, $k=4$, and $k=16$ curves when changing $\Omega_m$ / $\sigma_8$ by 5\%. The faint shaded regions indicate the range of $1\sigma$ variations at the fiducial cosmology.}
    \label{fig:knn_dr}
\end{figure}

\begin{figure}
    \centering
    \includegraphics[width=\linewidth]{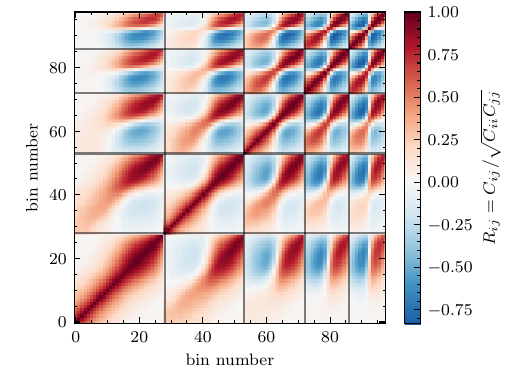}
    \caption{The DR-$k$NN correlation matrix. The sub-blocks along the diagonal correspond to $k = 1, 2, 4, 8, 16$.}
    \label{fig:knn_dr_cov}
\end{figure}

\begin{figure}
    \centering
    \includegraphics[width=0.9\linewidth]{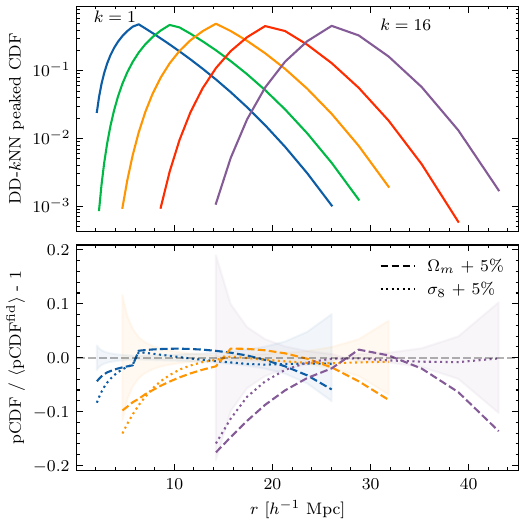}
    \caption{Same as Figure~\ref{fig:knn_dr}, but for the DD-$k$NNs.}
    \label{fig:knn_dd}
\end{figure}

\begin{figure}
    \centering
    \includegraphics[width=\linewidth]{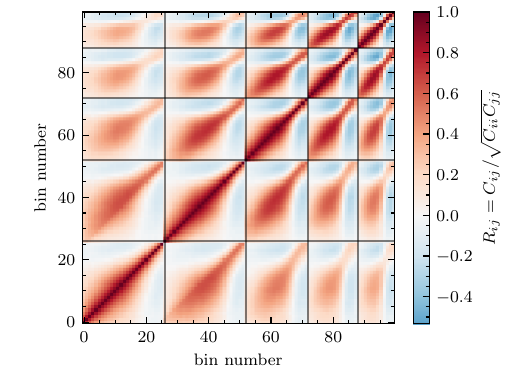}
    \caption{The DD-$k$NN correlation matrix. The sub-blocks along the diagonal correspond to $k = 1, 2, 4, 8, 16$.}
    \label{fig:knn_dd_cov}
\end{figure}

We show the average DR-$k$NNs for the fiducial simulations in Figure~\ref{fig:knn_dr} and their covariance in Figure~\ref{fig:knn_dr_cov}. Figures~\ref{fig:knn_dd} and \ref{fig:knn_dd_cov} show the same for the DD-$k$NNs.

\section{Fisher matrix formalism}
\label{sec:fisher}

The Fisher matrix is used ubiquitously in cosmology to estimate or forecast the constraints on cosmological parameters given some summary statistic and the expected measurement errors. In this work, we focus on using the Fisher matrix to compare the information content of different summary statistics rather than forecasting the cosmological constraints that could be obtained from a real experiment.

For some likelihood $\mathcal{L}$ with parameters of interest $\vb*{\theta}$, the Fisher matrix can be written as
\begin{equation}
    F_{\alpha\beta} = -\expval{\eval{\frac{\partial^2 \ln \mathcal{L}}{\partial\theta_\alpha\,\partial\theta_\beta}}_{\vb*{\theta}_0}}
\end{equation}
and characterizes the curvature of the likelihood around its peak $\vb*{\theta}_0$. This curvature around the peak directly limits how well we can constrain model parameters.

Assuming a Gaussian likelihood with a mean $\vb*{\mu}(\vb*{\theta})$ and a constant covariance matrix $C$, the Fisher matrix can be directly evaluated as
\begin{equation}
    \label{eq:fisher}
    F_{\alpha\beta} = \sum_{i,j} \frac{\partial \mu_i}{\partial \theta_\alpha} (C^{-1})_{ij} \frac{\partial \mu_j}{\partial \theta_\beta}.
\end{equation}

We explicitly confirm that our summary statistics roughly follow multivariate Gaussian distributions in Appendix~\ref{app:gauss}. We also ignore the dependence of the covariance matrix on cosmological parameters (as is often done in cosmological analyses) and simply assume a constant covariance matrix in our calculations of the Fisher matrix. As shown by \cite{Carron2013}, including this dependence can artificially inflate the estimated information content.

The error on a parameter $\theta_\alpha$, marginalizing over all other parameters, is then given by the Cram\'{e}r-Rao bound:
\begin{equation}
    \sigma_\alpha \ge \sqrt{(F^{-1})_{\alpha\alpha}}.
\end{equation}
We also define a figure-of-merit (FoM) to quantify how informative a given summary statistic is:
\begin{equation}
    \text{FoM} = \sqrt{\det F},
\end{equation}
which is proportional to the inverse of the volume enclosed by the parameter constraints. A larger FoM value indicates that the corresponding summary statistic is more informative about the parameters of interest.

\subsection{Covariances and derivatives}
For a given summary statistic, we calculate the covariance matrix using $N_\text{cov} = 5,000$ simulations at the fiducial parameter values. Because the covariance matrix is estimated from simulations, we also include the Hartlap correction factor \citep{Hartlap2007} by multiplying the estimated Fisher matrix from Equation~\ref{eq:fisher} by $(N_\text{cov} - n_b - 2) / (N_\text{cov} - 1)$, where $N_\text{cov}$ is the number of simulations used and $n_b$ is the number of bins in the data vector. This factor corrects for the fact that noise in the covariance matrix $C$ causes a biased estimate of $C^{-1}$ which inflates the Fisher matrix.

We estimate the derivatives of the summary statistics with respect to the cosmological parameters using finite differences. For the first five cosmological parameters (excluding $M_\nu$), we calculate the derivatives of the summary statistics using the centered second order finite difference formula. For $M_\nu$, we use the one-sided second order formula since the neutrino mass cannot be negative. Details on calculating these derivatives are given in \cite{Villaescusa-Navarro2020}. The finite difference derivative estimates are averaged over all $N_\text{deriv} = 500$ realizations available in the \Quijote suite to reduce cosmic variance and numerical noise.

\subsection{Convergence of Fisher forecasts}
Since both the covariance matrices and the summary statistic derivatives are being estimated from a finite number of simulations, noise can have a significant impact on the estimates of the Fisher matrices. 

As mentioned above we use the Hartlap factor to attempt to correct for noise in the estimated covariance matrices, but, mainly due to the large number of fiducial simulations used, we find that noise in the covariance matrices has negligible effects on the final estimates of the Fisher matrices (see Figure~\ref{fig:eigval_conv_cov}). 

We find a much larger effect due to the noise in the summary statistic derivatives. Any noise in the numerically estimated derivatives will bias the Fisher information high, leading to artificially low parameter constraints. This problem and potential solutions have been studied in detail recently by \cite{Coulton2023b} and \cite{Wilson2024}. \cite{Coulton2023b} suggest combining the traditional estimator (Equation~\ref{eq:fisher}) with one based on compressed summary statistics to get an estimate of the Fisher matrix that is less biased in the presence of noise. \cite{Wilson2024}, on the other hand, study this noise bias specifically in the context of the \Quijote simulations and demonstrate a method to calculate confidence intervals for Fisher forecasts. Both of these studies potentially indicate that the convergence of numerical derivatives in simulation-based Fisher matrices is a more significant issue than commonly thought in other recent studies.

Since we are not aiming to compute realistic forecasts, but want to compare the information content of the summary statistics as fairly as possible, we attempt to avoid these convergence problems altogether by determining which directions in parameter space are definitively converged. We do this by considering the eigendecomposition of the Fisher matrix. In general, the Fisher matrix for some cosmological parameters will have significant off-diagonal entries due to degeneracies between parameters. In the context of numerically estimating a Fisher matrix, these degeneracies can introduce significant issues: if even one of the parameter derivatives is noisy, inverting the matrix will propagate the noise through to the rest of the parameters, causing the marginalized uncertainties to be biased. We diagonalize the Fisher matrix to cleanly separate the different parameter directions that are being constrained. The corresponding eigenvalue for each eigenvector represents the Fisher information for that parameter combination, and since the eigenvectors are linearly independent, we can study the convergence of each eigenvalue individually.

We compute the Fisher matrix for each summary statistic for a range of values of $N_\text{deriv}$ between 10 and 500. For each value of $N_\text{deriv}$, we average over 100 different random selections of the derivative simulations. Considering a single direction in parameter space, if the corresponding estimate of the Fisher information is fully dominated by noise in the corresponding derivative, then we expect the magnitude of the Fisher information to go as
\begin{equation}
    F \propto 1/N_\text{deriv},
\end{equation}
where $N_\text{deriv}$ is the number of independent realizations used to estimate the derivative. This follows directly from Equation~\ref{eq:fisher} and the fact that the noise fluctuations in the derivative will go as $1/\sqrt{N_\text{deriv}}$. For each eigenvalue of the Fisher matrix, we fit a function of the form
\begin{equation}
    F(N_\text{deriv}) = F_\infty + F_\text{noise} / N_\text{deriv},
\end{equation}
where $F_\infty$ and $F_\text{noise}$ are free parameters that represent estimates of the true Fisher information and the amplitude of the noise contribution respectively. We find that this functional form provides a very good fit to the observed convergence behavior and it allows us to estimate the level of noise present in the different parameter directions. 

We show the full convergence properties of the Fisher matrices for the four considered summary statistics in Figure~\ref{fig:eigval_conv} and summarize the information about the best-constrained parameter directions in Table~\ref{tab:eigvecs}. We see that in all cases there are two directions for which the constraints are fully dominated by noise in the derivatives. As seen in Table~\ref{tab:eigvecs}, these unconstrained directions almost exactly correspond to $M_\nu$ and $h$, thus for all four summary statistics the constraints obtained for these parameters from the Fisher matrices are completely untrustworthy. On the other hand, for each summary statistic there are two parameter directions that are well constrained with estimated noise fractions $< 0.1$. 

Based on the fact that only two directions in parameter space have well converged constraints for all of the summary statistics considered, we choose to limit our main analysis to only two cosmological parameters: $\Omega_m$ and $\sigma_8$. In this reduced parameter space the Fisher matrices are well converged and we can be confident that the resulting parameter constraints are not affected by noise in the simulations. Because all of the Fisher matrices have a third direction that is in the process of converging, we choose to additionally show results for the expanded parameter space that includes $n_s$, under the caveat that these constraints will not be as well converged as those for the smaller parameter space.

\section{Results}
\label{sec:res}

\begin{figure*}
    \centering
    \includegraphics[width=0.48\linewidth]{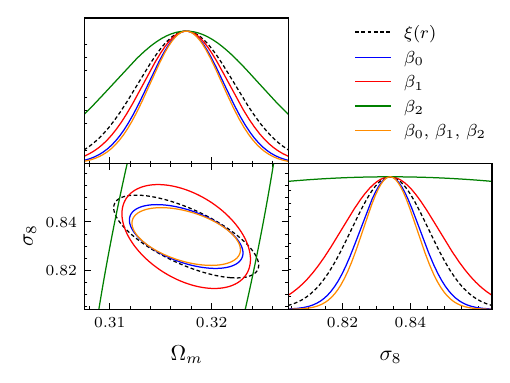}
    \includegraphics[width=0.48\linewidth]{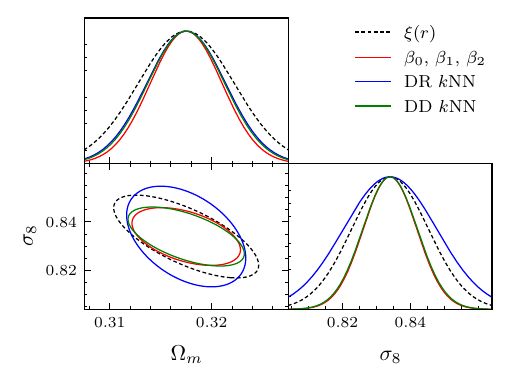}
    \caption{Constraints in the $\Omega_m$-$\sigma_8$ parameter space computed from the top $1.5\times10^5$ halos in real space. On the left we show the individual Betti curves and on the right we compare the combined Betti curves with the $k$NNs.}
    \label{fig:constr}
\end{figure*}

First, we compare the constraints that can be obtained from the individual Betti curves. In the left panel of Figure~\ref{fig:constr} we show the constraints obtained by considering each Betti curve individually. The first two Betti curves ($\beta_0$ and $\beta_1$), which track the number of connected components and loops respectively, have very similar constraining power on both $\Omega_m$ and $\sigma_8$, with $\beta_0$ providing slightly stronger constraints on both parameters. Perhaps surprisingly, $\beta_2$, which tracks the number of voids, provides weak constraints on $\Omega_m$ and almost no constraints on $\sigma_8$. In terms of the FoM values, we find that the combination of $\beta_0$ and $\beta_1$ provides 98\% of the information that is provided by the combination of all three Betti curves. 

We have also looked at the constraining power of the Euler characteristic\footnote{We thank Kwanit Gangopadhyay for suggesting this and note that \cite{Gangopadhyay2025} have shown a connection between $k$NNs and the Euler characteristic.}, defined as the alternating sum of Betti curves: $\chi = \beta_0 - \beta_1 + \beta_2.$
We find that the Euler characteristic provides constraints that are almost identical to the combination of $\beta_0$ and $\beta_1$, so it could provide an efficient compression of these summary statistics.

In the right panel of Figure~\ref{fig:constr}, we compare the parameter constraints obtained using the 2-point function, Betti curves, and $k$NN distributions. Here we see that all of the probes provide roughly similar constraints on $\Omega_m$, but the Betti curves and DD-$k$NNs provide significantly tighter constraints on $\sigma_8$. This provides evidence that the Betti curves and DD-$k$NNs are very efficient at extracting small-scale non-Gaussian information. 

\begin{figure*}
    \centering
    \includegraphics[width=0.48\linewidth]{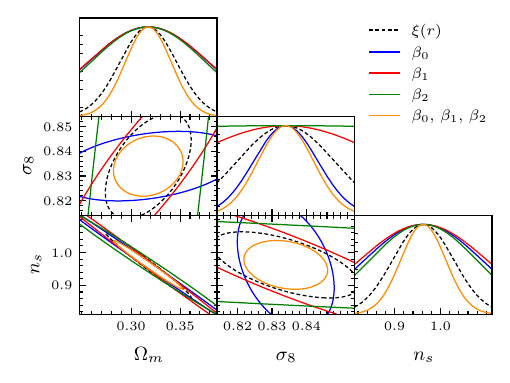}
    \includegraphics[width=0.48\linewidth]{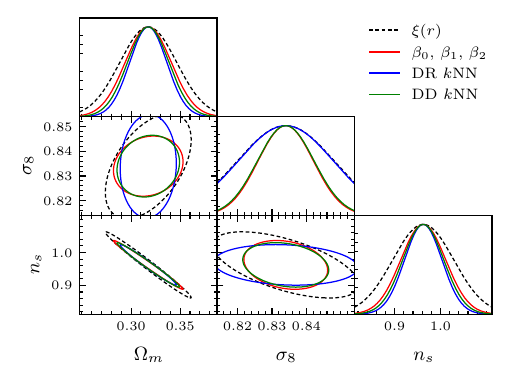}
    \caption{Same as Figure~\ref{fig:constr}, but for the $\Omega_m$-$\sigma_8$-$n_s$ parameter space.}
    \label{fig:constr3}
\end{figure*}

We show similar plots for the $\Omega_m$-$\sigma_8$-$n_s$ parameter space in Figure~\ref{fig:constr3}, where we see similar trends. Here, in contrast to the smaller parameter space, $\beta_2$ does add useful information, constraining $n_s$ better than the other Betti curves. Similarly, the DR-$k$NNs also do better in this larger parameter space by constraining $n_s$. As before, the Betti curves and DD-$k$NNs are significantly better at constraining $\sigma_8$. We speculate that this indicates that both the DD-$k$NNs and the combined Betti curves are primarily sensitive to high-density regions, while the DR-$k$NNs and $\beta_2$ are more sensitive to the low-density regions.

\begin{table}
    \centering
    \renewcommand{\arraystretch}{1.5}
    \begin{tabular}{lcc}
         & FoM ($\Omega_m$, $\sigma_8$) & FoM ($\Omega_m$, $\sigma_8$, $n_s$)$^\ast$ \\ \hline

         2-pt & 0.52 & 0.23 \\
         DR-$k$NN & 0.42 & 0.30 \\
         DD-$k$NN & 0.81 & 0.53 \\
         Betti & 0.81 & 0.49 \\ \hline

         2-pt + DR-$k$NN & 0.71 & 0.56 \\
         2-pt + DD-$k$NN & 0.82 & 0.58 \\
         2-pt + Betti & 0.89 & 0.62 \\
         DR-$k$NN + Betti & 0.83 & 0.71 \\
         DR-$k$NN + DD-$k$NN & 0.86 & 0.75 \\
         DD-$k$NN + Betti & 0.97 & 0.77 \\ \hline

         2-pt + DR-$k$NN + DD-$k$NN & 0.87 & 0.79 \\
         2-pt + DR-$k$NN + Betti & 0.90 & 0.82 \\
         2-pt + DD-$k$NN + Betti & 0.98 & 0.81 \\
         DR-$k$NN + DD-$k$NN + Betti & 0.99 & 0.97 \\ \hline
    \end{tabular}
    \caption{\textup{Amount of constraining power, defined as the FoM value normalized by the FoM for the combination of all probes, in the $\Omega_m$-$\sigma_8$ parameter space for each combination of probes. $^\ast$Warning: as noted in the text, the 3-parameter space is not as converged as the 2-parameter space, especially for the DR-$k$NNs.}}
    \label{tab:fom}
\end{table}

In order to characterize the amount of independent information coming from each probe, we list the FoM values (normalized by the FoM for the combination of all 4 probes) for all of the possible probe combinations in Table~\ref{tab:fom}. Starting with the individual probes, the DD-$k$NNs and the Betti curves provide very similar levels of constraining power that are significantly above those of the 2-point correlation function and the DR-$k$NNs in this analysis setup. The DR-$k$NNs do worse than the 2-point function in the reduced parameter space with weaker constraints on $\sigma_8$. A similar phenomenon was seen by \cite{Yuan2023}, where the 2-point function was more constraining on galaxy halo occupation parameters than the DR-$k$NNs, potentially due to the fact that the DR-$k$NNs primarily probe low-density regions. While in theory, both variants of the $k$NNs should strictly contain more information than the 2-point function for non-Gaussian fields \citep[see the derivations in][]{Banerjee2021}, this shows that the constraining power of $k$NNs may vary in practice due to a finite selection of $k$ values and/or scale cuts. Also, we have explicitly removed the tails of the $k$NN distributions in order to ensure that our summary statistics are Gaussian. If the variations of the $k$NN distributions are significantly non-Gaussian, then there could be significant small-scale information present in the tails which would necessitate more advanced likelihood characterization methods.

Combining two probes allows us to see that the Betti curves and $k$NNs do not contain completely independent information, since if two independent probes with equal constraining power are combined, we expect the FoM value to increase by a factor of $\sqrt{2}$. We see that combining the Betti curves with the 2-point function is actually more informative than combining the 2-point function with either of the $k$NNs. This provides evidence that the Betti curves are extracting more of the non-Gaussian information than the $k$NNs. We also note that the DD-$k$NN + Betti combination is very informative, almost approaching the constraining power of all four probes. Finally, combining three probes shows us that the 2-point function provides almost no new information once all of the $k$NNs and Betti curves are used.

In summary, we find that the Betti curves are very efficient at extracting non-Gaussian information and are highly complementary to the 2-point function. At the same time, $k$NNs provide complementary non-Gaussian information, but are not fully independent from the Betti curves. We speculate that this might indicate a deeper connection between these two probes.

\section{Discussion and Conclusions}
\label{sec:concl}

We have studied the cosmological information content of Betti curves, a summary statistic based on persistent homology, and $k$-nearest neighbor distance distributions. Previous studies have shown that these probes are both sensitive to non-Gaussian information in cosmological fields. Our comparison aims to quantify the degree of complementarity of these two probes and to look for potentially underlying connections between persistent homology and $k$NNs. Using halo catalogs from the \Quijote simulations we have shown that the combination of Betti curves and $k$NNs provides a very informative summary of non-Gaussian cosmological information. At the same time, these probes are not completely independent of each other, potentially indicating deeper connections. This hypothesis is also supported by the recent work by \cite{Gangopadhyay2025} that shows a connection between $k$NNs and Minkowski functionals, another topological summary.

This work is far from an exhaustive study of the cosmological information present in Betti curves. We note that the $\alpha$ complex is not the only possible choice of filtration one could make. The $\alpha$DTM$\ell$ complex has also been recently used for cosmological analyses \citep{Yip2024,Calles2024,Biagetti2021}. Compared to the $\alpha$ complex, the $\alpha$DTM$\ell$ complex is designed to be less sensitive to isolated points and is better at capturing cosmological voids. It is plausible that different variations of the filtration algorithm could be better adapted to different types of practical cosmological analyses.

While Betti curves might provide some complementary cosmological information compared to the $k$NNs, $k$NNs have some advantages that could make them more attractive for practical use. $k$NN statistics are significantly faster to compute than Betti curves and other similar topological summaries, due to the fact that the $k$-d tree data structure allows very efficient querying of nearest neighbor distances. Additionally, $k$NN statistics have a very natural decomposition in redshift space \citep{Yuan2023}, unlike Betti curves. Potentially, a similar decomposition could be implemented for the Betti curves using the formalism of multipersistence \citep[see for example,][]{2203.14289}. More work needs to be done to fully determine how effective Betti curves are at extracting non-Gaussian information from real galaxy catalogs.

\section*{Acknowledgements}
We thank Arka Banerjee and Kwanit Gangopadhyay for providing comments on our paper. 
This work was partially supported by the Center for AstroPhysical Surveys (CAPS) at the National Center for Supercomputing Applications (NCSA), University of Illinois Urbana-Champaign. A.O. acknowledges support from the Illinois Center for Advanced Studies of the Universe (ICASU)/CAPS/NCSA Graduate Fellowship. This work made use of the Illinois Campus Cluster, a computing resource that is operated by the Illinois Campus Cluster Program (ICCP) in conjunction with the National Center for Supercomputing Applications (NCSA) and which is supported by funds from the University of Illinois at Urbana-Champaign.

This analysis made use of many open-source software packages, including: \textsc{numpy} \citep{Harris2020}, \textsc{scipy}, \textsc{matplotlib} \citep{Hunter2007}, \textsc{IPython}, and \textsc{mpi4py} \citep{Dalcin2021}. Corner plots were generated using the \textsc{Getdist} package \citep{Lewis2019}. While writing this paper, we made extensive use of the Astrophysics Data System (ADS), the arXiv preprint server, and the \textsc{adstex} (\url{https://github.com/yymao/adstex}) package.

\section*{Data Availability}

The results of the \Quijote simulations are publicly available at \url{https://quijote-simulations.readthedocs.io/en/latest/index.html}. Other data can be made available upon reasonable request.

%%%%%%%%%%%%%%%%%%%% REFERENCES %%%%%%%%%%%%%%%%%%

\bibliographystyle{aasjournal}
\bibliography{references}

%%%%%%%%%%%%%%%%% APPENDICES %%%%%%%%%%%%%%%%%%%%%

\appendix

\section{Gaussianity of summary statistics}
\label{app:gauss}

Similar to \cite{Paillas2023}, we perform a qualitative test to ensure that the summary statistic roughly follow multivariate Gaussian distributions. For a given summary statistic, we use half of the 1,500 fiducial simulations to calculate the covariance matrix. We then use the estimated covariance matrix to calculate the $\chi^2$ statistic for each of the simulations in the other half.
\begin{equation}
    \chi^2_i = (\vb*{d}_i - \vb*{\mu})^T C^{-1} (\vb*{d}_i - \vb*{\mu})
\end{equation}
We have included the Hartlap correction factor \citep{Hartlap2007} when using the estimated covariance matrix to calculate the inverse covariance matrix $C^{-1}$

\begin{figure*}
    \centering
    \includegraphics[width=\linewidth]{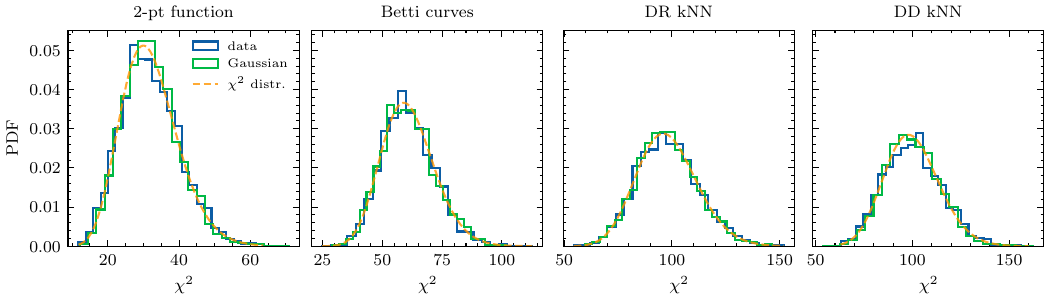}
    \caption{The distribution of $\chi^2$ values for each of the considered summary statistics. The blue histograms show the distributions of $\chi^2$ values calculated from the data vectors. The orange curves show the expected distributions from theory, and the green histograms show realizations sampled from multivariate Gaussian distributions with the same means and covariances as the data vectors.}
    \label{fig:chi2}
\end{figure*}

If the summary statistic is drawn from a multivariate Gaussian, then the calculated $\chi^2$ values should follow a $\chi^2$ distribution with $N_b$ degrees of freedom. As shown in Figure~\ref{fig:chi2}, we find that all of the used summary statistics match this expected behavior fairly well and we see no significant evidence for non-Gaussian behavior.

\section{Convergence of parameter constraints}
\label{app:conv}

\begin{figure*}
    \centering
    \includegraphics[width=\linewidth]{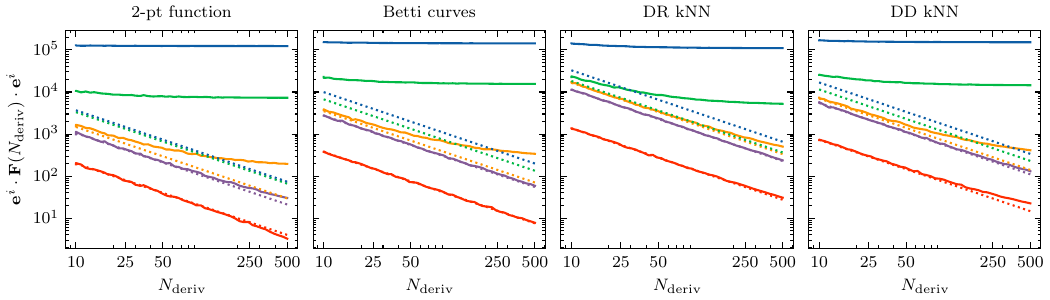}
    \caption{Convergence of Fisher matrix eigenvalues with respect to the number of derivative simulations. The dotted lines indicate the estimated noise level for each eigenvalue.}
    \label{fig:eigval_conv}
\end{figure*}

We show the convergence of the Fisher matrix eigenvalues with respect to the number of derivative simulations $N_\text{deriv}$ in Figure~\ref{fig:eigval_conv} and list the respective eigenvectors and estimated noise fractions in Table~\ref{tab:eigvecs}.

\begin{figure*}
    \centering\includegraphics[width=\linewidth]{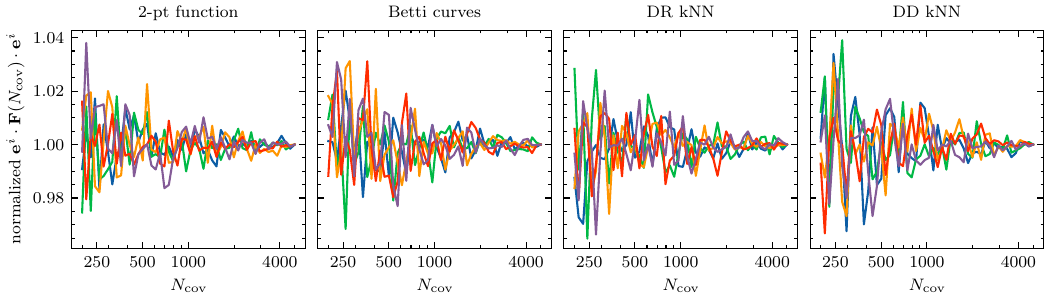}
    \caption{Convergence of the Fisher matrix eigenvalues with respect to the number of covariance simulations.}
    \label{fig:eigval_conv_cov}
\end{figure*}

In Figure~\ref{fig:eigval_conv_cov}, we show the convergence of eigenvalues with respect to the number of simulations used to estimate the covariance matrices of each considered probe. The parameter constraints are very well converged with respect to $N_{\text{cov}}$, the number of simulations used to calculate the fiducial covariance. 

\begin{table}[h]
    \centering
    \renewcommand{\arraystretch}{2}
    \begin{tabular}{M{1.2cm}M{3.5cm}M{1.5cm}M{1cm}}
        Summary \newline statistic & Eigenvectors \newline ($\Omega_m$, $\sigma_8$, $h$, $n_s$, $M_\nu$) & Eigenvalues & Noise fraction \\ \hline
        
        2-point function \newline \newline
        & \textbf{-0.83}, -0.26, -0.29, -0.39, ~0.01 \newline
          -0.26, \textbf{~0.96}, -0.11, ~0.01, ~0.05 \newline 
          ~0.45, ~0.11, -0.26, \textbf{-0.84}, -0.09 \newline
          -0.16, ~0.06, \textbf{~0.91}, -0.37, ~0.04 \newline
          ~0.06, -0.03, -0.05, -0.06, \textbf{~0.99}
        & $1.24\times10^5$ \newline $7.30\times10^3$ \newline
          $1.95\times10^2$ \newline $3.03\times10^1$ \newline
          $3.25\times10^0$
        & 0.001 \newline 0.009 \newline 0.15 \newline 0.71 
          \newline 1
        \\ \hline
               
        Betti curves \newline \newline
        & \textbf{-0.84}, -0.20, -0.30, -0.41, ~0.01 \newline
          -0.18, \textbf{~0.98}, -0.07, -0.05, ~0.05 \newline
          ~0.43, ~0.03, -0.01, \textbf{-0.90}, -0.03 \newline
          -0.27, ~0.01, \textbf{~0.95}, -0.14, ~0.04 \newline
          ~0.04, -0.05, -0.04, -0.01, \textbf{~1.00}
        & $1.44\times10^5$ \newline $1.56\times10^4$ \newline
          $3.39\times10^2$ \newline $5.88\times10^1$ \newline
          $7.77\times10^0$
        & 0.001 \newline 0.008 \newline 0.21 \newline 0.94
          \newline 0.97
        \\ \hline

        DR-$k$NN \newline \newline
        & \textbf{-0.86}, -0.11, -0.30, -0.40, ~0.02 \newline
          -0.11, \textbf{~0.99}, -0.04, ~0.00, ~0.04 \newline
          ~0.45, ~0.04, -0.14, \textbf{-0.88}, ~0.00 \newline
          -0.22, ~0.02, \textbf{~0.94}, -0.26, -0.01 \newline
          ~0.02, -0.04, ~0.02, ~0.01, \textbf{~1.00}
        & $1.10\times10^5$ \newline $5.25\times10^3$ \newline
          $5.03\times10^2$ \newline $2.37\times10^2$ \newline
          $3.08\times10^1$
        & 0.006 \newline 0.070 \newline 0.65 \newline 0.95
          \newline 0.89
        \\ \hline

        DD-$k$NN \newline \newline
        & \textbf{-0.83}, -0.25, -0.29, -0.40, ~0.01 \newline
          -0.23, \textbf{~0.97}, -0.10, -0.07, ~0.03 \newline
          ~0.43, ~0.04, -0.03, \textbf{~0.90}, -0.06 \newline
          -0.27, ~0.03, \textbf{~0.95}, -0.15, -0.08 \newline
          ~0.02, -0.03, ~0.08, -0.06, \textbf{~0.99}
        & $1.53\times10^5$ \newline $1.45\times10^4$ \newline
          $4.12\times10^2$ \newline $1.30\times10^2$ \newline
          $2.25\times10^1$
        & 0.002 \newline 0.016 \newline 0.34 \newline 0.84
          \newline 0.65
        \\ \hline
    \end{tabular}
    \caption{\textup{Eigenvectors and corresponding eigenvalues of the Fisher matrices for the four considered summary statistics. For each eigenvector, the largest component is bold to indicate with which parameter the vector most lines up with and we list the estimated fraction of noise present in the corresponding eigenvalue due to non-converged derivatives.}}
    \label{tab:eigvecs}
\end{table}

\bigskip

%%%%%%%%%%%%%%%%%%%%%%%%%%%%%%%%%%%%%%%%%%%%%%%%%%

\end{document}